\documentclass[onefignum,onetabnum]{siamart171218}

\usepackage{lipsum}
\usepackage{amsfonts}
\usepackage{graphicx}
\usepackage{epstopdf}
\usepackage{algorithmic}
\ifpdf
  \DeclareGraphicsExtensions{.eps,.pdf,.png,.jpg}
\else
  \DeclareGraphicsExtensions{.eps}
\fi

\newsiamremark{remark}{Remark}
\newsiamremark{hypothesis}{Hypothesis}
\crefname{hypothesis}{Hypothesis}{Hypotheses}
\newsiamthm{claim}{Claim}

\headers{Oblique wave injection}{F. P\'erez and M. Grech }

\title{Oblique-incidence, arbitrary-profile wave injection for electromagnetic simulations%
}

\author{F. P\'erez\thanks{Laboratoire pour l'Utilisation des Lasers Intenses, CNRS, \'Ecole Polytechnique, CEA, Universit\'e Paris-Saclay, Sorbonne Universit\'e, F-91128, Palaiseau Cedex, France}
\and
M. Grech\footnotemark[1]}

\usepackage{amsopn}

\usepackage{enumitem}
\usepackage{xspace}
\newcommand{\Smilei}{{\sc Smilei}\xspace}

\begin{document}

\maketitle

\begin{abstract}
In an electromagnetic code, a wave can be injected in the simulation domain
by prescribing an oscillating field profile at the domain boundary.
The process is straightforward when the field profile has a known analytical expression
(typically, paraxial Gaussian beams). However, if the field profile is
known at some other plane, but not at the boundary (typically, non-paraxial
beams), some pre-processing is needed to calculate the field profile after propagation
back to the boundary. We present a parallel numerical technique for this propagation
between an arbitrary tilted plane and a given boundary of the simulation domain,
implemented in the Maxwell-Vlasov particle-in-cell code \Smilei.
\end{abstract}

\section{Introduction}

Electromagnetic (EM) codes are popular tools in various fields of physics,
from nonlinear photonics to laser-matter and laser-plasma physics.
In such codes, an EM wave can be introduced in the simulation domain by two means.
The EM wave can be imposed as an initial condition, i.e. prescribing the fields
throughout the domain at the initial time of simulation, of course ensuring
that these fields satisfy the Poisson and zero-magnetic-divergence equations.
This approach requires a simulation domain large enough to contain the whole
EM wave. Furthermore, the knowledge of the full spatial profile at a given time
may be challenging to obtain.
To remove these constraints, a second technique, which we consider in the present
article, consists in imposing
the EM wave as a time-varying boundary condition. This second approach only
requires the boundary surface to be large enough, and the knowledge of the EM
field spatio-temporal profile is only required at the boundary.

Although the present article is relevant to any EM code or Maxwell solver,
we illustrate the
proposed method with Particle-In-Cell (PIC) simulations \cite{birdsall1985}
using the open-source PIC code \Smilei \cite{derouillat2018}.
The PIC method simulates the self-consistent evolution of both the field and
particle distribution of a plasma.
It is widely used, from astrophysical studies \cite{grassi2016, martins2009}
to ultra-intense laser-plasma interaction \cite{tajima1979, thaury2010}.
In \Smilei, Maxwell's equations are solved
using the Finite-Difference-Time-Domain approach \cite{taflove2005,nuter2014},
and EM waves can be injected/absorbed using the \textit{Silver-M\"uller}
boundary conditions \cite{barucq1997}.
The latter allow for EM wave injection by prescribing the transverse
magnetic field profiles at a boundary of the simulation domain.

Planar waves and paraxial Gaussian beams have a direct analytical formulation of the
EM field in the whole space. Specifying their field as a function of time at
one given boundary is thus trivial. However, other profiles do not have an
analytical representation, or at least not in the whole space. Often, they are known in a
given plane which does not correspond to a simulation box boundary. This is typical for
experimental profiles or theoretical non-paraxial beams.
In that case, specifying the field profiles at a given boundary is more involved.
The present article describes a method to facilitate this process.

A recent work by Thiele \textit{et al.} \cite{thiele2016} details a technique to
pre-process an EM wave profile specified at a given plane (parallel to, but not at the boundary) in order
to propagate it backward and obtain the field profiles at the boundary.
This approach is largely based on the Angular Spectrum Method (ASM)
used in other domains, such as acoustics \cite{stepanishen1982, clement2000} and digital
holography \cite{matsushima2003}. It consists in applying a propagation factor to the fields
in the spatial frequency domain, thus relies heavily on Fourier transforms.
The theory by Thiele \textit{et al.} extends this principle to temporal frequencies,
consequently allowing to prescribe a temporal profile to the EM wave.
This is obviously a strong requirement, in PIC codes, for modeling ultra-short
(femtosecond) laser pulses.

In the present article, we combine the work by Thiele \textit{et al.} and that of
Matsushima \textit{et al.} \cite{matsushima2003} in order to pre-process ultra-short-pulse
EM waves prescribed at an arbitrary, oblique plane inside the simulation domain.
The paper is organized as follows.
Section \ref{parallel-propagation} summarizes the theory by Thiele \textit{et al.}, which we
complete by that of  Matsushima \textit{et al.} in Section \ref{tilted-propagation}. The
numerical technique deployed in \Smilei is presented in Section \ref{numerics}, and
examples in two and three dimensions are given in Section \ref{results}.
Finally, our conclusions are given in Section \ref{conclusion}. 

\section{Propagation between parallel planes}\label{parallel-propagation}

\begin{figure}[htbp]
\begin{center}
\includegraphics{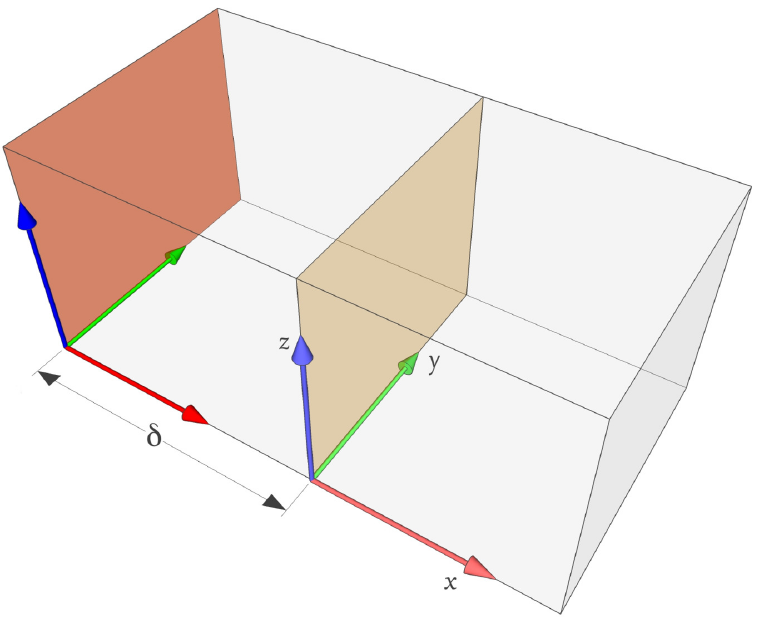}
\caption{Box model of a simulation where the EM wave is prescribed in a parallel plane
at a distance $\delta$ from a boundary.}
\label{geometry1}
\end{center}
\end{figure}

The ASM theory can be summarized in a simple manner (note that the following
discussions relate to three-dimensional simulations, but can be directly applied to
two-dimensional simulations by discarding the $z$ axis.). 
It is valid for any scalar field $A$
satisfying a wave equation:

\begin{equation}\label{wave-equation}
    c^2 \Delta A(x,y,z,t) = \partial_t^2 A(x,y,z,t)
\end{equation}
\noindent
where $x$, $y$, $z$ and $t$ are the space and time coordinates, and $c$ is the wave velocity.
In our situation, the field $A$ may be any component of the EM field
(or of the EM vector potential), and $c$
is the speed of light in vacuum, as the propagation is only considered without plasma. We
study specifically a propagation along the $x$ axis, between two planes $x=-\delta$ and
$x=0$, as illustrated in figure \ref{geometry1}.

The three-dimensional Fourier transform of equation \ref{wave-equation} for the variables $y$, $z$ and
$t$ gives:

\begin{equation}\label{Fourier-wave-equation}
    (\partial_x^2 + k_x^2) \hat A(x,k_y,k_z,\omega) = 0
\end{equation}
\noindent
where $k_y$, $k_z$ and $\omega$ are the respective conjugate variables, in the frequency
domain, of $y$, $z$ and $t$; and $k_x(k_y,k_z,\omega) \equiv
\sqrt{\omega^2/c^2-k_y^2-k_z^2}$. Equation \eqref{Fourier-wave-equation} has general
solutions proportional to $\exp(-i k_x x)$ for waves propagating towards positive $x$.
This means that, if the profile $A$ is known at $x=0$, the profile at $x=-\delta$ is
obtained after multiplying $\hat A$ by $\exp(i k_x \delta)$:
\begin{equation}\label{general-parallel-propagation}
    \hat A(-\delta,k_y,k_z,\omega) = \exp(i k_x \delta) \hat A(0,k_y,k_z,\omega)
\end{equation}
\noindent

Note that the function $k_x(k_y,k_z,\omega)$ assumes pure imaginary values
where $k_y^2 + k_z^2 > \omega^2/c^2$. As those correspond to evanescent waves, they should not
contribute to the propagation, and are simply removed from the calculation. In
other terms, equation \eqref{general-parallel-propagation} should be replaced by:
\begin{equation}
    \hat A(-\delta,k_y,k_z,\omega) = P_\delta(k_y, k_z, \omega)\, \hat
A(0,k_y,k_z,\omega)
\end{equation}
\noindent
where we have introduced the propagation factor:
\begin{equation}\label{propagation-term}
  \begin{array}{l}
    P_\delta(k_y, k_z, \omega) =
    \,\left\{
    \begin{array}{ll}
        \exp\left(i k_x(k_y,k_z,\omega) \delta\right) & \textrm{if
}\omega^2/c^2>k_y^2+k_z^2\\
        0 & \textrm{otherwise.}
    \end{array}
    \right.
  \end{array}
\end{equation}

To recover the field profile $A(-\delta,y,z,t)$ in real space, a
three-dimensional inverse Fourier transform would be sufficient. However, storing all
values of the $(y,z,t)$ profile might consume too
much time and disk space. Instead, as suggested in Ref. \cite{thiele2016}, only a
two-dimensional inverse Fourier transform on $k_y$ and $k_z$ may be carried out. This
results in a $\tilde A(-\delta,y,z,\omega)$ profile, where $\omega$ still corresponds to the
temporal Fourier modes. If necessary, only a few of these modes 
(the most intense ones) can be kept to ensure a reasonable disk-space usage.

In the end, the full $A(-\delta,y,z,t)$ profile is calculated during the actual PIC
simulation, summing over the different $\omega$ according to:
\begin{equation}\label{reconstruction}
  \begin{array}{l}
    A(-\delta,y,z,t) = 
    f(y,z,t) \, \sum_\omega \left| \tilde A(-\delta,y,z,\omega) \right| \sin\left(\omega t +
\phi(-\delta,y,z,\omega)\right)
  \end{array}
\end{equation}
\noindent
where $\phi$ is the complex argument of $\tilde A$ and $f(y,z,t)$ is an additional
profile, defined by the user. This optional profile $f$ provides some extra control over
the temporal reconstruction of the wave: as a finite number of temporal modes may be kept, the
reconstructed wave is periodic, thus spurious repetitions of the EM pulse may occur
later during the simulation. This custom function $f$ may be used to remove those unwanted
repetitions.


\section{Propagation between tilted planes} \label{tilted-propagation}

\begin{figure}[htbp]
\begin{center}
\includegraphics{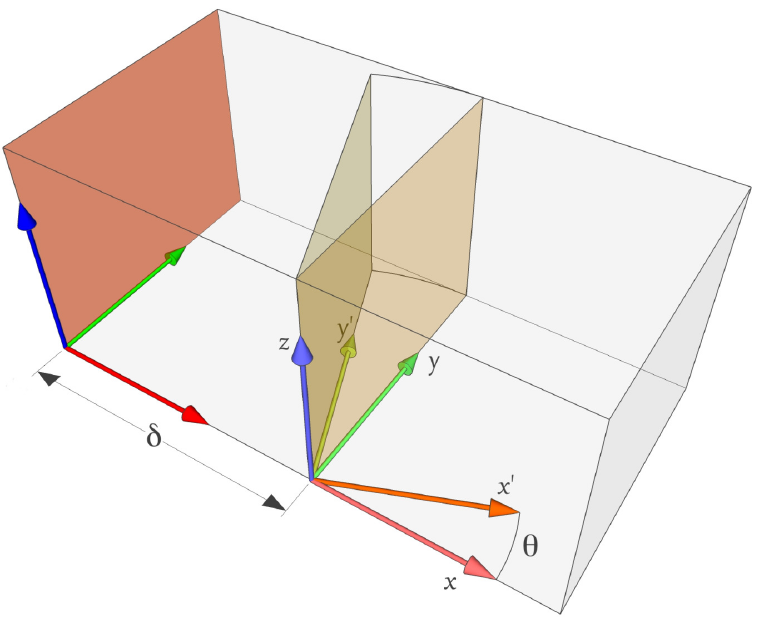}
\caption{Box model of a simulation where the EM wave is prescribed in a oblique plane.}
\label{geometry2}
\end{center}
\end{figure}

In the context of image reconstitution for digital holography, 
Matsushima \textit{et al.} \cite{matsushima2003} proposed an extension of the ASM to handle the propagation between
two non-parallel planes, as illustrated in figure \ref{geometry2}.
In this section, we summarize a version of that theory that
follows the previous section's formulation.

The rotation of the wave is handled in the Fourier space, that is working directly on the field $\hat
A(x,k_y,k_z)$. In this section, the argument $\omega$ is not
explicitly written as it does not change the reasoning.

We consider, for simplicity, a rotation around the axis of $k_z$.
Denoting rotated quantities with a ``prime'' symbol, the rotation of the wave vector by an
angle $\theta$ reads:
\begin{equation}\label{rot-transform}
    \begin{array}{rcl}
    k_x^\prime & = & k_x \cos\theta  + k_y \sin\theta \\
    k_y^\prime & = & -k_x \sin\theta  + k_y \cos\theta \\
    k_z^\prime & = & k_z
    \end{array}
\end{equation}
\noindent
where $k_x^\prime \equiv \sqrt{\omega^2/c^2-k_y^{\prime 2}-k_z^{\prime 2}}$.

To obtain an expression of the rotated profile
$\hat A^\prime(x^\prime,k_y^\prime,k_z^\prime)$, let us first
apply a propagator term $\exp(i k_x x)$ in the Fourier transform:
\begin{equation}\label{transform1}
\begin{array}{rcl}
A(x,y,z) & = & \iint\hat A(x,k_y,k_z) \exp\left(i k_y y + i k_z z\right) \mathrm d k_y \mathrm d k_z\\
         & = & \iint\hat A(x=0,k_y,k_z) \exp\left(\vec k\cdot\vec r\right) \mathrm d k_y \mathrm d k_z
\end{array}
\end{equation}
where $\vec k = (k_x, k_y, k_z)$ and $\vec r = (x,y,z)$.
Note that this equation is also true in the rotated frame:
\begin{equation}\label{transform2}
\begin{array}{rcl}
A^\prime(x^\prime,y^\prime,z^\prime) & = \iint\hat A^\prime(x^\prime=0,k_y^\prime,k_z^\prime)
\exp\left(\vec{k^\prime}\cdot\vec{r^\prime}\right)
 \mathrm d k_y^\prime \mathrm d k_z^\prime
\end{array}
\end{equation}
Knowing that rotation preserves the scalar product
$\vec k\cdot\vec r$, we may operate a change of variable
$(k_y,k_z)\rightarrow(k_y^\prime,k_z^\prime)$ in equation
\eqref{transform1}:
\begin{equation}\label{transform3}
\begin{array}{rcl}
A(x,y,z) & = & \iint{ a(k_y^\prime,k_z^\prime)
 \exp\left(\vec{k^\prime}\cdot\vec{r^\prime}\right)
 \left|\cos\theta-\frac{k_y^\prime}{k_x^\prime}\sin\theta\right|
 \mathrm d k_y^\prime \mathrm d k_z^\prime}
\end{array}
\end{equation}
where we have introduced $a(k_y^\prime,k_z^\prime)
= \hat A(x=0,k_y,k_z)$.
As the original profile is equal to the rotated one in rotated coordinates,
i.e. $A(x,y,z)=A^\prime(x^\prime,y^\prime,z^\prime)$, identifying
the integrands in equations \eqref{transform2} and \eqref{transform3}
results in:
\begin{equation}\label{transform4}
\begin{array}{rcl}
\hat A(x=0,k_y,k_z) = \hat A^\prime(x^\prime=0,k_y^\prime,k_z^\prime)
\left|\cos\theta-\frac{k_y^\prime}{k_x^\prime}\sin\theta\right|^{-1}
\end{array}
\end{equation}
This last equation constitutes the calculation to be carried out
to obtain the rotated wave at $x=0$. More precisely, the user
prescribes $A^\prime(x^\prime=0, y^\prime, z^\prime)$ which is
converted to $\hat A^\prime(x^\prime=0, k_y^\prime, k_z^\prime)$
using a double Fourier transform. Then, equation \eqref{transform4}
consists in interpolating values on the rotated Fourier space,
and applying the factor to obtain $\hat A(x=0,k_y,k_z)$.
After this rotation, one may apply the same propagation factor as
in equation \eqref{propagation-term} in order to combine both
a rotation and a translation.
Follows an inverse Fourier transform, identical to the description of the previous
section, to recover the profile in real space.

This technique applies to any scalar field, and by extension, to any component of the EM field.
In \Smilei, the boundary conditions only require the knowledge of the magnetic field profiles transverse to the boundary.
As an example, for the $x=0$ boundary, one needs to prescribe the components $B_y$ and $B_z$,
the other components being naturally obtained by solving Maxwell's equations. 
Hence, in the rotated frame, the user needs to define the profiles $B^\prime_{y^\prime}(x^\prime=0,y^\prime,z^\prime)$ and
$B^\prime_{z^\prime}(x^\prime=0,y^\prime,z^\prime)$, i.e.
the magnetic field projected to $y^\prime$  and $z^\prime$ only.
The rotation and propagation back to the boundary $x=0$ leads to the fields
$B_{y^\prime}(x=0,y,z)$ and $B_{z^\prime}(x=0,y,z)$, which lie in the
$x=0$ plane, but remain the components projected to $y^\prime$  and $z^\prime$.
To recover the components along the simulation directions $y$ and $z$, a simple projection is applied:
\begin{equation}\label{projection}
  \begin{array}{l}
    B_{y} = B_{y^\prime}\cos\theta\\
    B_{z} = B_{z^\prime}
  \end{array}
\end{equation}


\section{Numerical implementation} \label{numerics}

The theory described in the previous sections presents a few numerical obstacles requiring
careful treatment, especially to enable a parallel treatment on a large number of processors. Let us
first summarize the general numerical process.
In the particular case of \Smilei, this involves arrays of complex
numbers of initial size $(N_y,N_z,N_t)$ corresponding, by default, to the number of cells
$N_y$ and $N_z$ of the spatial mesh and the number of timesteps $N_t$ required for the PIC
simulation. The final array size is reduced to $(N_y,N_z,M_t)$, with $M_t<N_t$, should one
keep only some of the temporal Fourier modes (as described in Section \ref{parallel-propagation}).
The process is outlined as follows.

\begin{enumerate}
\item The EM wave profile $A^\prime$ defined by the user is evaluated for all $(y^\prime,z^\prime,t^\prime)$
coordinates.
\item Its Fourier transform $\hat A^\prime$ is computed along the three axes.
\item For each $\omega$, the total spectral energy is computed, and only those $\omega$
with the highest magnitude (according to some user-defined criterion) are kept.
\item An interpolation method transforms $\hat A^\prime$ into $\hat A$ corresponding to a
rotation in frequency space, see equation \eqref{transform4}.
\item The array $\hat A$ is multiplied by the propagation factor of equation
\eqref{propagation-term}.
\item The inverse Fourier transform is computed along $k_y$ and $k_z$ only.
\item The resulting array is stored in a file. The file is read later by each process, in
order to reconstruct the wave, at each timestep, according to equation
\eqref{reconstruction}
\end{enumerate}

Steps 2 and 6 make use of an implementation of the Fast Fourier Transform (FFT) algorithm.
However, to enable multi-parallel computation, the arrays must be split among
$N_\textrm{proc}$ processors. There are several approaches to perform a parallel FFT
(see Ref. \cite{richard} for a review).
In \Smilei, we have chosen to split the initial array
$A^\prime$ in equal parts along its first dimension, $y$ (note that, before step 1, the array size is
extended, if necessary, to a multiple of $N_\textrm{proc}$). As a consequence, each
processor owns an array of size $(n_y,N_z,N_t)$, where $n_y=N_y/N_\textrm{proc}$. Step 2
begins by applying the FFT algorithm to the last two dimensions $z$ and $t$. In order to
transform the first dimension $y$, some data re-organization is necessary. We employ the
Message Passing Interface (MPI) protocol to communicate parts of the arrays between
processors, so that each processor finally owns a portion of the global array
corresponding to a slab along the second dimension $z$, of size $(N_y,n_z,N_t)$, where
$n_z=N_z/N_\textrm{proc}$. This allows the computation of the FFT along $y$, thus
providing the total Fourier transform $\hat A^\prime$. Note that, in this whole process,
the decomposition of the array between processors has changed.

In step 6, the calculation is almost the same, but in reverse order. First, the inverse
FFT is computed along $y$. Then, the array decomposition between processors is reversed
again, using MPI, so that they each own a portion of size $(n_y,N_z,M_t)$. Finally, the
inverse FFT is computed over the axis $z$.

Step 4 requires some interpolation algorithm that maps $\hat
A^\prime(k_y^\prime,k_z,\omega)$ to $\hat A(k_y,k_z,\omega)$. In computing terms, for each
point at location $(k_y,k_z,\omega)$ of the resulting array $\hat A$, we determine the
location $(k_y^\prime,k_z,\omega)$ in the initial array $\hat A^\prime$ from where the
value is copied, using equation \eqref{rot-transform}. As this location may not fall
exactly on a array point, we must interpolate between two consecutive $k_y^\prime$.
Importantly, the complex numbers given by a typical FFT of a laser profile present a
rapidly-varying argument, often close to, or faster than the grid resolution.
Consequently, one must not interpolate linearly between two complex numbers to avoid
cancellation of two numbers with opposite phases. Instead, in \Smilei, we considered that
the magnitude and argument are both physically significant: the former represents the
\textit{weight} assigned to each mode, while the latter corresponds to a delay in space
and/or time. This consideration supports a separate interpolation for the magnitude and
the argument of these complex numbers, which is done in \Smilei. One additional precaution
is necessary for the interpolation of the argument: as the global phase is supposed to
vary smoothly across the array, we ensure that two consecutive arguments are always in the
same order (e.g., the second larger than the first, adding $2\pi$ to the second when
necessary).

\section{Examples}\label{results}

The overall method described in the previous sections has been implemented in \Smilei in
both two- and three-dimensional Cartesian geometries.

\begin{figure}[htbp]
\begin{center}
\includegraphics[width=0.9\textwidth]{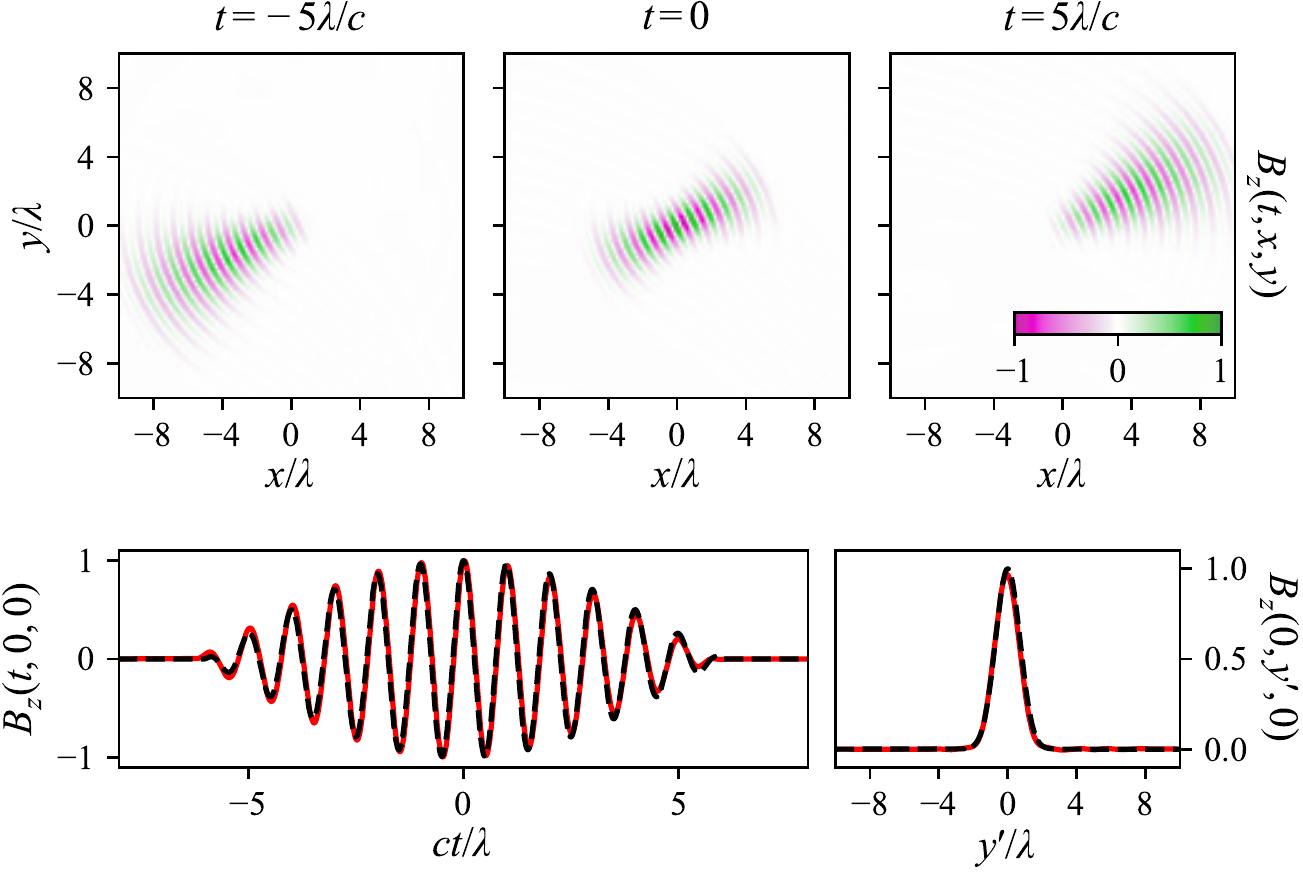}
\caption{Two-dimensional simulation of a tightly-focused beam prescribed in a tilted plane.
Top panels: Magnetic field $B_z$ at three different times.
Bottom panels: Comparison of the obtained (red lines) and requested (dashed black lines)
magnetic field profiles, as a function of time (left) at $x=y=0$, and as a function
of $y^\prime$ (right) at $t=0$.
}
\label{tst2d}
\end{center}
\end{figure}

Let us first present a two-dimensional simulation of a tightly-focused laser pulse
with an angle of incidence of 25$^\circ$, a wavelength $\lambda$ and a linear
polarization in the simulation plane ($x,y$).
The full simulation domain extends from $x=-16\lambda$ to $16\lambda$ and $y=-96\lambda$ to $96\lambda$.
To inject this laser, we prescribed a $B_z$ magnetic field profile along a line
$y'$ tilted by 25$^\circ$ with respect to the $x=-16\lambda$ boundary of the simulation
domain. The intensity profile is Gaussian in space
and has a $\cos^2$ shape in time, the focus being located in the middle of
the simulation box ($x=0$, $y=0$, $y'=0$) and the time $t=0$ denoting the time
at which the laser field is maximal:
\begin{eqnarray}
B_z(t,y') = a_0\,\exp\!\left(-\frac{y'^2}{w^2}\right)\,
\cos\!\left(\frac{\pi}{2}\frac{t}{\tau}\right)\,
\Pi\!\left(\frac{t}{\tau}\right)\,\cos(\omega_0t)\,
\end{eqnarray}
with $\omega_0=2\pi\,c/\lambda$ the laser angular frequency,
$\Pi(t)=1$ for $-1<t<1$ and 0 otherwise, $a_0=1$ the 
field amplitude in arbitrary units,
$w=\lambda$ the waist, and $\tau=6\lambda/c$ the duration.
The spatial and temporal resolutions were set to $\Delta x = \Delta y = \lambda/32$
and $c\Delta t=0.95 \Delta x/\sqrt{2}$, respectively.
For this simulation, only the 128 most intense modes were kept.
The simulation results are reported in Fig.~\ref{tst2d}.
The top panels show both the spatial and temporal aspects of the laser
pulse propagation.
The bottom panels prove the excellent quantitative agreement between
the simulated field evolution and the prescribed profiles.
 
\begin{figure}[htbp]
\begin{center}
\includegraphics[width=0.9\textwidth]{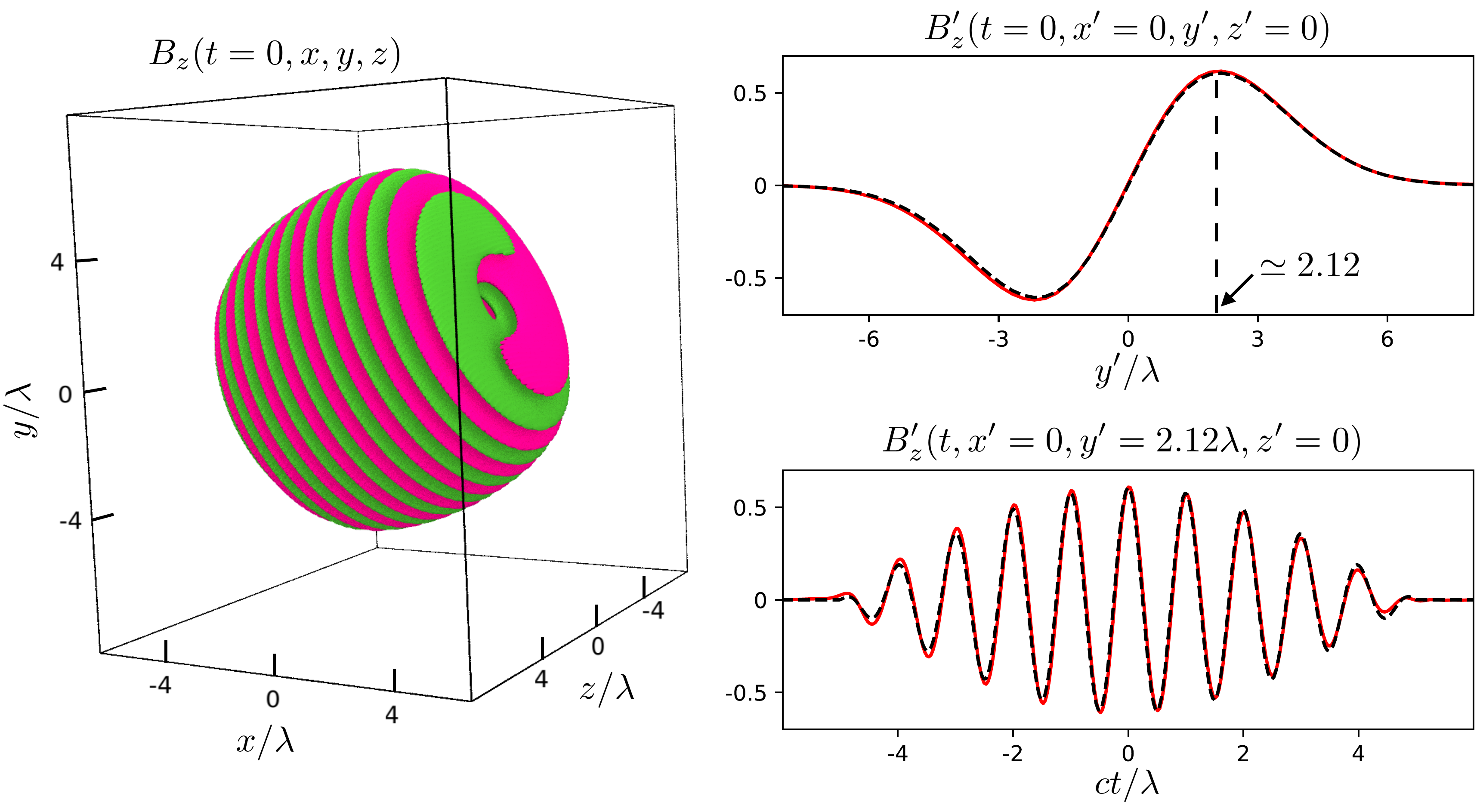}
\caption{Three-dimensional simulation of an oblique-incidence
	Laguerre-Gauss beam. 
	Left: 3D rendering of magnetic field isocontours at $B_z=\pm a_0/10$.
	Top right: spatial profile in the tilted plane at $t=0$.
	Bottom right: temporal profile at $x^\prime=z^\prime=0$, $y^\prime=2.12\lambda$.
	Red and dashed black lines represent obtained and requested
	profiles, respectively.
}
\label{tst3d}
\end{center}
\end{figure}

To illustrate our method in a three-dimensional simulation,
we prescribe, in a $(y^\prime, z)$ plane titled by $25^\circ$,
a Laguerre-Gauss beam (mode $1,0$) with a $\cos^2$ temporal shape as:
\begin{eqnarray}
B_z(t,y',z) = a_0\,\sqrt{2}\,\frac{r^\prime}{w}\,\exp\!\left(-\frac{r^{\prime 2}}{w^2}\right)\,
\cos\!\left(\frac{\pi}{2}\frac{t}{\tau}\right)\,
\Pi\!\left(\frac{t}{\tau}\right)\,\cos\left(\omega_0t-\phi^\prime\right)
\end{eqnarray}
where $r^\prime=\sqrt{y^{\prime 2}+z^2}$, $\phi^\prime=\arctan(z/y^\prime)$,
$w=3\lambda$, and $\tau=5 \lambda/c$.
The simulation box extends
from $x=-8\lambda$ to $8\lambda$,
from $y=-24\lambda$ to $24\lambda$, and
from $z=-16\lambda$ to $16\lambda$.
The spatial and temporal resolutions are $\Delta x=\Delta y=\lambda/16$ and $c\Delta t=0.95 \Delta x/\sqrt{3}$, respectively.
Fig. \ref{tst3d} illustrates the propagated laser pulse
in a three-dimensional rendering of $B_z$ isocontours.
The right-hand-side panels show the excellent quantitative agreement
between the obtained and requested field profiles.

\section{Conclusion}\label{conclusion}

In summary, the present article reviews and combines two related theories for the
propagation of waves between two planes. They both extend the ASM, the first for temporal
profiling, and the second for propagation between tilted planes. We have combined both
approaches and described the numerical implementation in a parallel processing environment
using the open-source PIC code \Smilei. It will be applied to several situations relevant
to high-intensity laser-plasma interaction: tightly-focused, or spatially-chirped laser
pulses, and waves featuring orbital angular momentum.

\section*{Acknowledgments}
The authors thank Rachel Nuter for useful discussions and comparisons with her PIC code,
and the \Smilei development team for technical support. 
This work was granted access to HPC resources from GENCI-TGCC (Grant No. 2017-x2016057678).

\end{document}